\documentclass[12pt]{article}
\newcommand{\Real}[1]{\Re {\it e}(#1 )}
\newcommand{\Imag}[1]{\Im {\it m}(#1 )}
\newcommand{\xrm}[1]{{\textstyle \mbox{\rm #1}}}
\newcommand{\fnd}[2]{\frac{\textstyle #1}{\textstyle #2}}
\newcommand{\abs}[1]{\left| #1\right|}
\newcommand{\bm}[1]{\mbox{\boldmath $#1$}}
\begin{document}
\title{\bf The light scalar mesons within quark models}
\author{{\large Eef van Beveren{\normalsize $^{\; a}$},
George Rupp{\normalsize $^{\; b}$}},\\ [.3cm]
{\large Nicholas Petropoulos{\normalsize $^{\; a}$} and
Frieder Kleefeld{\normalsize $^{\; b}$}}\\ [.5cm]
{\normalsize\it $^{a\;}$Centro de F\'{\i}sica Te\'{o}rica, Departamento de
F\'{\i}sica,}\\ {\normalsize\it Universidade, P3004-516 Coimbra, Portugal,}
\\ [.1cm]
{\small (eef@teor.fis.uc.pt and nicholas@teor.fis.uc.pt)}\\ [.3cm]
{\normalsize\it $^{b\;}$Centro de F\'{\i}sica das Interac\c{c}\~{o}es
Fundamentais, Instituto Superior T\'{e}cnico,}\\ {\normalsize\it
Edif\'{\i}cio Ci\^{e}ncia,
P1049-001 Lisboa Codex, Portugal,}\\ [.1cm]
{\small (george@ajax.ist.utl.pt and kleefeld@cfif.ist.utl.pt)}\\ [1cm]
{\normalsize Contribution to the}\\ {\normalsize\bf
Second International Workshop on Hadron Physics,}\\ {\normalsize\bf
Effective Theories of Low Energy QCD}\\
{\normalsize at the}\\
{\normalsize Centro de F\'{\i}sica Te\'{o}rica da Universidade de Coimbra}
{\normalsize 25-29 September, 2002}
}
\maketitle

\begin{abstract}
Low-energy meson-meson scattering data are a powerful testing ground for
quark models.
Here, we describe the behaviour at threshold of $S$-wave scattering-matrix
singularities.

The majority of the full scattering-matrix mesonic poles
stem from an underlying confinement spectrum.
However, the light scalar mesons $K_{0}^{\ast}(830)$, $a_{0}(980)$,
$f_{0}$(400--1200), and $f_{0}(980)$ do not,
but instead originate in $^{3}P_{0}$-barrier semi-bound states.
We show that the behaviour of the corresponding poles is
identical at threshold.

In passing, the light-meson sector is given a firm basis.
\end{abstract}
\clearpage

{\bf Introduction.}
It is generally understood that, once the meson sector of strong interactions
is fully and consistently described, not too many complications are expected
upon including the baryons as well.
However, as things stand, important steps have yet to be taken towards
a simple quantitative theory for the description of mesons and their
interactions \cite{HEPPH0112205}.
Preferably, this should be a one-parameter theory which, in the limit of free
quarks and gluons and some non-abelian interaction, approaches QCD.
Here, we pay attention to the unification of all flavours.

For a lowest-order approximation of strong interactions,
confinement models may be constructed.
Their usefulness can be measured by the models' achievements
when adjusting their parameters to experiment\cite{PRD32p189}.
Observed spectra may be interpreted in terms of quark-antiquark
or more complicated systems \cite{HEPPH0206263}.
But in a full theory with quarks and mesons, one can study strong interactions
through meson-meson scattering.
Consequently, one needs more than confinement only.
For further refinements of strong-interaction models, one should
compare the models' predictions directly to experimental
scattering cross sections and phase shifts when available
\cite{Cargese75p305,PRD21p203,PRD27p1527,ZPC30p615,NPB266p451,PRL76p1575,HEPPH0110081,HEPPH0203255}.

Here we will discuss the eternally disputed
\cite{HEPPH0204205,HEPPH0201171}
low-lying nonet of $S$-wave poles in meson-meson scattering cross sections,
within a four-parameter model \cite{ZPC30p615}.
It should thereby be noted that the model's parameters
are fitted to the $J^{P}=1^{-}$ $c\bar{c}$ and $b\bar{b}$ spectra,
as well as to $P$-wave meson-meson scattering data \cite{PRD27p1527}.

First, let us outline the motivation for our work.
For the interaction in the vicinity of a resonance in meson-meson
scattering, one may consider quark-exchange or quark-pair-creation processes,
giving rise to an intermediate $q\bar{q}$ system.
When the intermediate $q\bar{q}$ system is close enough to a genuine
bound state of confinement, then the system will resonate,
resulting in a resonance in meson-meson scattering.
Another picture for the same phenomenon is to consider self-energy
contributions from virtual meson loops.
Either picture describes the same physical situation,
namely a mesonic resonance or bound state \cite{PLB509p81},
but in a rather different way.
Our aim is to merge both pictures in one model.
\vspace{0.2cm}

{\bf Flavour-independent confinement.}
Let us assume that the spectrum of mesonic quark-antiquark systems
can be described by flavour-independent harmonic-oscillator confinement.
Then, for each pair of flavours, an infinite set of mesons exists with all
possible spin, angular, and radial excitations.
But, unfortunately, for most flavour pairs only a few angular and even fewer
radial recurrencies are known \cite{PRD66p010001}.
When we do not distinguish up and down, but just refer to non-strange ($n$)
quarks and, moreover, ignore the existence of top quarks, then we dispose
of four different flavours: $n$, $s$, $c$, and $b$.
These can be combined into ten different flavour pairs, each of which may come
in two different spin states: 0 or 1.
This gives rise to, in principle, twenty different meson spectra.
With some 150 known mesons, this means 7.5 angular plus radial
excitations on average, per flavour pair.  This is much less than e.g.\ the
known excitations of the positronium spectrum.
No wonder that it requires some imagination to guess economic strategies for
the description of mesons.

\begin{figure}[htbp]
\begin{center}
\begin{picture}(135,150)(-35,-2)
\put(-5.0,125.40){\makebox(0,0)[rc]{GeV}}
\put(-8.0,  6.00){\makebox(0,0)[rc]{ 0.5}}
\put(-8.0, 36.00){\makebox(0,0)[rc]{ 1.0}}
\put(-8.0, 66.00){\makebox(0,0)[rc]{ 1.5}}
\put(-8.0, 96.00){\makebox(0,0)[rc]{ 2.0}}
\put( 14.40, 16.27){\makebox(0,0)[tc]{\footnotesize $\rho$}}
\put( 45.80, 37.20){\makebox(0,0)[lc]{1S}}
\put( 45.80, 60.00){\makebox(0,0)[lc]{2S 1D}}
\put( 45.80, 82.80){\makebox(0,0)[lc]{3S 2D}}
\put( 45.80,105.60){\makebox(0,0)[lc]{4S 3D}}
\put( 45.80,128.40){\makebox(0,0)[lc]{5S 4D}}
\end{picture}
\begin{picture}(135,150)(-35,-5)
\put(-5.0,121.20){\makebox(0,0)[rc]{GeV}}
\put(-8.0, 12.00){\makebox(0,0)[rc]{ 3.0}}
\put(-8.0, 42.00){\makebox(0,0)[rc]{ 3.5}}
\put(-8.0, 72.00){\makebox(0,0)[rc]{ 4.0}}
\put(-8.0,102.00){\makebox(0,0)[rc]{ 4.5}}
\put( 14.40, 11.81){\makebox(0,0)[tc]{\footnotesize $J/\Psi$}}
\put( 45.80, 32.40){\makebox(0,0)[lc]{1S}}
\put( 45.80, 55.20){\makebox(0,0)[lc]{2S 1D}}
\put( 45.80, 78.00){\makebox(0,0)[lc]{3S 2D}}
\put( 45.80,100.80){\makebox(0,0)[lc]{4S 3D}}
\put( 45.80,123.60){\makebox(0,0)[lc]{5S 4D}}
\end{picture}
\begin{picture}(125,150)(-35,8)
\put(-5.0,133.20){\makebox(0,0)[rc]{GeV}}
\put(-8.0, 30.00){\makebox(0,0)[rc]{ 9.5}}
\put(-8.0, 60.00){\makebox(0,0)[rc]{10.0}}
\put(-8.0, 90.00){\makebox(0,0)[rc]{10.5}}
\put(-8.0,120.00){\makebox(0,0)[rc]{11.0}}
\put( 14.40, 21.62){\makebox(0,0)[tc]{\footnotesize $\Upsilon$}}
\put( 45.80, 46.20){\makebox(0,0)[lc]{1S}}
\put( 45.80, 69.00){\makebox(0,0)[lc]{2S 1D}}
\put( 45.80, 91.80){\makebox(0,0)[lc]{3S 2D}}
\put( 45.80,114.60){\makebox(0,0)[lc]{4S 3D}}
\put( 45.80,137.40){\makebox(0,0)[lc]{5S 4D}}
\end{picture}
\end{center}
\normalsize
\caption[]{Non-strange, charmonium, and bottomonium $J^{PC}=1^{--}$ states
compared to the corresponding states from a harmonic-oscillator spectrum.
The level spacing for the oscillator equals 0.38 GeV.}
\label{vectors}
\end{figure}
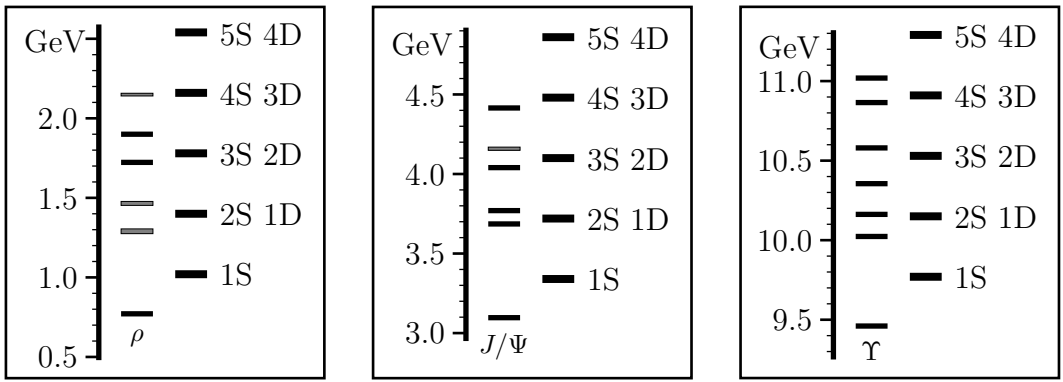

As one may verify from the latest Review of Particle Physics
\cite{PRD66p010001},
vector states are more numerous than any other type of mesonic resonances,
hence better known.
Consequently, in order to structure a model we begin with the vector mesons,
carrying quantum numbers $J^{P}=1^{-}$.
In figure (\ref{vectors}) we compare the measured $n\bar{n}$, $c\bar{c}$, and
$b\bar{b}$ vector states with the possible states of the harmonic oscillator.
Most of the data are taken from Ref.~\cite{PRD66p010001}.
The $\rho$(1290) signal has been reported in
Refs.~\cite{SLACPUB5606,NPPS21p105,JINRP286682,NCA49p207,NPB76p375}.
The $\Upsilon (1D)$ is our interpretation of the $D$ state which has been
observed in Ref.~\cite{HEPEX0207060}.

The charmonium vector states, shown in figure (\ref{vectors}),
bear many similarities with the two-particle harmonic oscillator:
a ground state in $S$ wave, and higher radial excitations that are
{\it almost} \/degenerate with the $D$-wave states.
Also, except for the ground state, the level spacings are roughly equal.
In Ref.~\cite{HEPPH0201006} the mechanism is discussed which turns
the oscillator spectrum into the charmonium spectrum.

For the $\rho$ and $\Upsilon$ vector states, also shown in figure
(\ref{vectors}), we see a very similar pattern:
the $S$-$D$ splittings are slightly larger, while the $\rho$(770) ground
state of the $\rho$ spectrum and the $\Upsilon(1S)$ ground state of the
$\Upsilon$ spectrum also come out far below the corresponding oscillator
ground states.
From figure (\ref{vectors}) one may moreover conclude that there is
not much reason to separate the light-quark sector from the heavy quarks.
Below, we discuss the mechanism which turns the oscillator states into
the $\rho$ and $\Upsilon$ resonances \cite{PRD27p1527}.

What we learn from the above comparison is that confinement is flavour
independent.
Hence, confinement should be described by flavour-independent dynamics.
A non-relativistic Schr\"{o}dinger equation with
flavour-mass-dependent harmonic-oscillator \cite{PRD27p1527} is just
a perfect example of such dynamics.
It casts quark confinement in the form of a one-parameter model.
This parameter is the oscillator frequency $\omega$, which comes out at about
0.19 GeV for the data.
\vspace{0.2cm}

{\bf Mechanism for more structure.}
At this point we dispose of a beautiful one-parameter model for mesons,
which has a particularly simple spectrum for each of the flavour and spin
excitations, given by

\begin{equation}
M\left( f,\bar{f};\ell ,n)\right)\; =\;
\omega\;\left( 2n+\ell +\frac{3}{2}\right)\;
+\; m_{f}\; +\; m_{\bar{f}}
\;\;\; .
\label{HOspectrum}
\end{equation}

\noindent
Here, $f$ and $\bar{f}$ represent the flavours of respectively the quark and
the antiquark, $m_{f}$ and $m_{\bar{f}}$ their respective masses,
$\ell$ and $n$ their relative angular momentum and radial excitation.

Let us study some details of formula (\ref{HOspectrum}) in the following.
The vector-meson states have unit total angular momentum, $J=1$, and
unit $q\bar{q}$ total spin, $s=1$.
Hence, since the parity of vector-meson states equals $P=-1$,
their orbital angular momentum can be $\ell =0$ ($S$ wave) or
$\ell =2$ ($D$ wave).
From formula (\ref{HOspectrum}) we then understand that, within
harmonic-oscillator confinement, the vector-meson states
with ($n$, $\ell =2$) are degenerate with the vector-meson states
with ($n+1$, $\ell =0$), as shown in figure (\ref{vectors}).
For other flavour and spin excitations similar results emerge.
One obtains a very regular, equally spaced spectrum of quark-antiquark states.
However, by comparing the experimental meson spectrum to the
theoretical harmonic oscillator spectrum (\ref{HOspectrum}), we readily
see that our simple model by far does not agree with the data.

In order to cure this disagreement, we may study other potentials,
which generate spectra that agree better with the data, and even modify,
whenever necessary, their parameters for different flavour sectors.
But here we prefer to stick to our one-parameter model for all flavours.

When we ignore the electroweak interactions, then the mesons of our model are
permanently stable quark-antiquark systems.
For the great majority of mesons, this picture badly conflicts with
observation.
Strong interactions are not just confinement, but also hadronic decay, and
elastic and inelastic scattering of hadrons.
Our one-parameter model just describes confinement.
All other strong phenomena must still be included.

Hadronic decay is quantitatively well described by the phenomenon of
quark-pair creation \cite{CERNREPTH401/412},
i.e., the creation of a valence quark and a valence antiquark
out of the vacuum.
Quark pairs are supposed to be created and annihilated all the time inside
the realm of a hadron.
Only once in a while such pair develops into a valence quark pair, which allows
the hadron to decay into other hadrons.
We will represent the probability for that process to occur
by one parameter, $\lambda$, which, in the spirit of flavour symmetry,
we assume to be constant for all flavours.
In practice, we only consider the creation of non-strange and strange
valence quark pairs, since the thresholds for charm and bottom are much
higher \cite{PRD65p034001},
far above the few mesonic resonances which we want to describe.

Both $\lambda$ and the oscillator frequency $\omega$ should be related to
the fundamental parameter $\alpha$ of QCD. However, having no knowledge about
such relation, we accept them as free parameters.
Nevertheless, they are chosen independent of the flavour pair which
constitutes the meson under consideration.
This way we guarantee flavour independence of our model's results, as
dictated by QCD, and reconfirmed by experiment \cite{PRD59p012002}.

\begin{figure}[htbp]
\begin{center}
\small
\begin{picture}(226.77,141.73)(-46.00,-17.60)
\put(181.45,-5.08){\makebox(0,0)[tr]{interquark distance}}
\put(-5.08,114.56){\makebox(0,0)[tr]{$V_{t}$}}
\end{picture}
\end{center}

\normalsize
\caption[]{Form of the transition potential $V_{t}$
in the coordinate representation}
\label{Vtrans}
\end{figure}
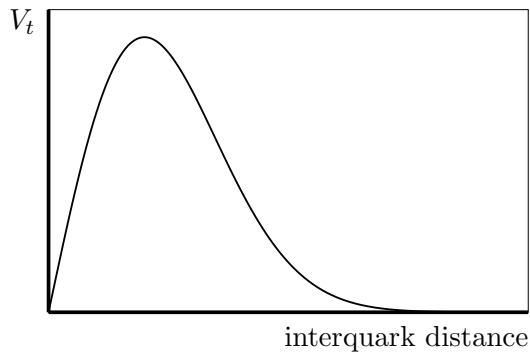

We suppose that quark-pair creation takes place in the interior of a hadron,
neither at very large (confinement), nor at very small (asymptotic
freedom) interquark distances.
In the coordinate representation, these considerations get translated
into a potential $V_{t}$ of the form as depicted in figure (\ref{Vtrans}).
This transition potential enables the communication between a meson and
its decay products.

The principle of flavour independence of strong interactions demands that
the transition potential cannot be a function of the pure interquark
distance $r$, but has to be scaled by the reduced mass $\mu$ of the flavours of
the system, {\it i.e.}

\begin{equation}
V_{t}\; =\; V_{t}\left(\sqrt{\mu}\, r\right)
\;\;\; .
\label{vtrans}
\end{equation}

\noindent
The model's results are then to lowest order flavour independent.
Differences, which also can be observed in the data, stem from the higher
orders and kinematics.

We have chosen the form of $V_{t}$ such that it depends on two,
again flavour-independent, parameters.
This completes the description of the model, which, besides the four
constituent quark masses, has four model parameters.
\vspace{0.2cm}

{\bf Non-exotic meson-meson scattering.}
The model under discussion is taylor-made for describing
non-exotic meson-meson scattering.
In particular, for processes of the form

\begin{equation}
\xrm{meson }A\; +\;\xrm{meson }B\;\longrightarrow\;
q\bar{q}\;\longrightarrow\;
\xrm{meson }C\; +\;\xrm{meson }D
\;\;\; ,
\label{ABtoCD}
\end{equation}

\noindent
where it is understood that all flavours are such that the $q\bar{q}$ system
couples to the $AB$ two-meson system by light-quark-pair annihilation and
to the $CD$ two-meson system by light-quark-pair creation.
Processes with more than two mesons in the final state are supposed
to come from subsequent decays of the mesons in the initial $CD$ two-meson
final state, and hence to be of less importance for the properties we want
to study.

In the interaction region, where our model applies, we have a finite
probability to find an $AB$ two-meson system, or a $CD$ two-meson system,
or a $q\bar{q}$ mesonic system.
The latter is usually far from its confinement eigenstates and thus highly
unstable.
The $CD$ pair represents any of the unlimited number of possible
two-meson final states.
In Ref.~\cite{ZPC21p291} it is described how the two-meson final states
can be limited to a finite number of possibilities.
Additionally, we only take the lowest-lying pseudoscalar, $\pi$, $K$, $\eta$,
or $\eta '$, and vector mesons $\rho$, $K^{\ast}$, $\omega$, or $\phi$,
as final-state particles.
Nevertheless, the number of final-state channels comes out mostly somewhere
between ten and twenty.
However, in Ref.~\cite{HEPEX0106077} it is shown that,
when one takes relatively small errors of up to some 50 MeV for granted,
then one may even obtain part of the results for low energies by just
including the lowest-lying, or the most important, final-state channel.

The above-discussed parameter $\lambda$ represents the overall
three-meson-vertex coupling of the two-meson channels to the $q\bar{q}$ channel
of expression (\ref{ABtoCD}).
Relative couplings for the various three-meson vertices that may occur
can be determined by a technique described in Ref.~\cite{ZPC21p291}.
In some cases one also has various possible $q\bar{q}$ channels, as
the isosinglets, $n\bar{n}$, $s\bar{s}$, $\ldots$ mix with one another.

The probability to find a certain two-particle state in the interaction region
can be determined from its wave function.
Consequently, the model can be formulated in terms of a multichannel system of
coupled-channel equations for the various wave functions.
Details may be found in Ref.~\cite{PRD27p1527}.
By a technique which has been described in Ref.~\cite{LNP211p182}, one can
analytically solve the coupled-channel
equations and determine the scattering matrix $S(\sqrt{s})$ as a function
of the total centre-of-mass energy.
Subsequently one may study elastic cross sections, or phase shifts, but
also inelasticities and wave functions.
Moreover, the solutions of the coupled-channel equations may be analytically
continued to complex values of $\sqrt{s}$, which allows to study the
pole structure of the scattering matrix in the complex-energy plane.
\clearpage

{\bf Bound states and resonances.}
Each of the two-meson final-state channels has a minimum value for the
energy at which the two mesons can be formed, the threshold energy,
which is given by the sum of the two meson masses.
When the total energy $\sqrt{s}$ of the coupled-channel system is above
the threshold of a particular channel, one says that the channel is open.
It means that scattering is possible for that channel.
Nevertheless, it is legitimate to study the solutions of the set of
coupled-channel equations for energies below the thresholds,
i.e., when channels are closed.
In particular, below the lowest threshold, where all channels are closed and
no scattering is possible, one may obtain the analytic
continuation in $\sqrt{s}$ of the scattering matrix.

Singularities in $S(\sqrt{s})$ for real values of $\sqrt{s}$ below the lowest
threshold, represent permanently bound states, when calculated in the correct
Riemann sheet (i.e., with positive imaginary momenta $k$ in all channels).
The wave functions corresponding to those singularities have for all channels
contributions that rapidly vanish at large interparticle distances.
They represent the stable mesons, like the $J/\Psi$, which cannot decay
strongly.
For ground states of pseudoscalar and vector mesons, one finds the poles
shifted to mass values which are far below the corresponding oscillator
ground states.
The sizes of the shifts are roughly proportional to $\lambda^{2}$, and may
have values of several hundreds of MeV.
The higher excitations of pseudoscalar and vector mesons shift much less.
For this reason, one may start from harmonic-oscillator confinement and yet
end up with realistic meson spectra \cite{HEPPH0201006}.
We achieve this for all mesons with one fixed value for each of the four model
parameters \cite{PRD27p1527}.

The fact that all channels contribute to the wave function of stable mesons
implies that stable mesons, too, have two-meson components, not just
$q\bar{q}$. This observation has important consequences for electromagnetic and
weak transitions of stable mesons \cite{PRD44p2803}.

Above the lowest threshold no bound states can exsist.
Accordingly, one does not find singularities in the scattering
matrix on the real $\sqrt{s}$ axis.
The poles which are found in that region of the complex $\sqrt{s}$
plane have real and imaginary parts.
When a pole is encountered in the lower half of the complex $\sqrt{s}$ plane,
and moreover in the right Riemann sheet and not too far from
the real axis, then one may observe a nearby enhancement in the elastic cross
sections of all open channels.
These structures represent the resonances that show up in meson-meson
scattering, as e.g.\ the $\rho$ meson in $\pi\pi$ scattering, and the
$K^{\ast}$ meson in $K\pi$ scattering.

In the quark-exchange picture, we obtain a resonance in the particular
partial-wave meson-meson-scattering cross section which matches
the quantum numbers of the intermediate $q\bar{q}$ system of process
(\ref{ABtoCD}).
Such a phenomenon may be described by scattering phase shifts of the
Breit-Wigner \cite{PR49p519} form

\begin{equation}
\xrm{cotg}\left(\delta_{\ell}(s)\right)\;\approx\;
\fnd{E_{R}-\sqrt{s}}{\Gamma_{R}/2}
\;\;\; ,
\label{cotgdR}
\end{equation}

\noindent
where $E_{R}$ and $\Gamma_{R}$ represent the central
invariant meson-meson mass and the resonance width, respectively.

However, formula (\ref{cotgdR}) is a good approximation for the
scattering cross section only when the resonance shape is not very much
distorted and the width of the resonance is small.
Moreover, the intermediate state in such a process is essentially
a constituent $q\bar{q}$ configuration that is part of a
confinement spectrum (also referred to as bare or intrinsic state),
and hence may resonate in one of the eigenstates.
This implies that the colliding mesons scatter off
the whole $q\bar{q}$ confinement spectrum of radial, and possibly
also angular excitations, not just off one single state \cite{NC14p951}.
Consequently, a full expression for the phase shifts of formula
(\ref{cotgdR}) should contain all possible eigenstates of such a
spectrum as long as quantum numbers are respected.
Let us denote the eigenvalues of the relevant part of the spectrum
by $E_{n}$ ($n=0$, $1$, $2$, $\dots$),
and the corresponding eigenstates by ${\cal F}_{n}$.
Then, following the procedure outlined
in Ref.~\cite{HEPEX0106077},
we may write for the partial-wave phase shifts the
more general expression

\begin{equation}
\xrm{cotg}\left(\delta (s)\right)\; =\;
\left[ I(s)\;\sum_{n=0}^{\infty}\fnd{
\abs{{\cal F}_{n}}^{2}}{\sqrt{s}-E_{n}}\right]^{-1}\;
\left[ R(s)\;\sum_{n=0}^{\infty}\fnd{
\abs{{\cal F}_{n}}^{2}}{\sqrt{s}-E_{n}}\; -\; 1\right]
\;\;\; .
\label{cotgdS}
\end{equation}

\noindent
In $R(s)$ and $I(s)$ we have absorbed the kinematical factors
and details of two-meson scattering, and
moreover the three-meson vertices.
The details of formula (\ref{cotgdS}) can be found in
Ref.~\cite{HEPEX0106077}.

For an approximate description of a specific resonance, and in
the rather hypothetical case that the three-meson vertices have small
coupling constants, one may single out, from the sum over all
confinement states, one particular state (say number $N$),
the eigenvalue of which is nearest to the
invariant meson-meson mass close to the resonance.
Then, for total invariant meson-meson masses $\sqrt{s}$ in the vicinity
of $E_{N}$, one finds the approximation

\begin{equation}
\xrm{cotg}\left(\delta (s)\right)\;\approx\;
\fnd{\left[ E_{N}\; +\; R(s)\;\abs{{\cal F}_{N}}^{2}\right]\; -\;\sqrt{s}}
{I(s)\;\abs{{\cal F}_{N}}^{2}}
\;\;\; .
\label{cotgdSs}
\end{equation}

\noindent
Formula (\ref{cotgdSs}) is indeed of the form (\ref{cotgdR}),
with the central resonance position and width given by

\begin{equation}
E_{R}\;\approx\; E_{N}\; +\; R(s)\;\abs{{\cal F}_{N}}^{2}
\;\;\;\;\;\xrm{and}\;\;\;\;\;
\Gamma_{R}\;\approx\; 2I(s)\;\abs{{\cal F}_{N}}^{2}
\;\;\; .
\label{ERGR}
\end{equation}

\noindent
In experiment one observes the influence of the nearest bound state of
the confinement spectrum, as in classical resonating systems.
Nevertheless, formula (\ref{cotgdSs}) is only a good approximation
when the three-meson couplings are small.
Since the coupling of the meson-meson system to quark exchange
is strong, the influence of the higher- and lower-lying excitations is
not negligible.

In data analyses one usually represents a resonance in the elastic cross
section by a Breit-Wigner structure,
associated with a singularity in the complex $\sqrt{s}$ plane.
In our model resonances are not well represented by Breit-Wigner structures,
due to non-perturbative effects.
A good example is the $\rho '$(1250).
This resonance is omitted from the meson tables of the Review of
Particle Physics.
Nevertheless, it is reported in data analyses as a clear signal
\cite{SLACPUB5606,NPPS21p105,NPB76p375}.
In our model it comes out as a very tiny structure around 1.26 GeV in
coupled $\pi\pi$, $K\bar{K}$, $\eta_{n}\rho$, $\pi\omega$, $KK^{\ast}$,
$\rho\rho$, $K^{\ast}K^{\ast}$ scattering in the tail of the $\rho$(770)
resonance \cite{PRD27p1527}.
Now, since in the data analyses quoted in \cite{PRD66p010001} one uses
Breit-Wigner structures, it is no surprise that for this tiny effect, which
moreover has a rather large width, no evidence is found.
At the position of the $\rho '$(1470), which in Ref.~\cite{PRD66p010001} is
claimed to be the first radial $\rho$ excitation, we find a $D$ resonance,
which has the same quantum numbers as the $S$ state.
Naturally, we may know which resonance is dominantly $S$ and which is
dominantly $D$, since we can determine the wave functions in our model.
From cross sections alone one cannot easily distinguish \cite{ZPC31p77}
between the two.

In the other hypothetical limit, namely of very large couplings, we obtain
for the phase shift the expression

\begin{equation}
\xrm{cotg}\left(\delta (s)\right)\;\approx\;
\fnd{R(s)}{I(s)}
\;\;\; ,
\label{cotgdSl}
\end{equation}

\noindent
which describes scattering off an infinitely hard cavity.

The physical values of the couplings come out somewhere in between
the two hypothetical cases.
Most resonances and bound states can be classified as stemming from a
specific confinement state \cite{PLB413p137,HEPPH0204328}.
However, some structures in the scattering cross section stem from the
cavity which is formed by quark exchange or pair creation
\cite{HEPEX0106077}.
The most notable of such states are the low-lying resonances
observed in $S$-wave pseudoscalar-pseudoscalar scattering
\cite{HEPEX0012009,HEPEX0110052,HEPEX0204018,HEPPH0110156}.

From the above discussion one may conclude
that, to lowest order, the mass of a meson follows from
the quark-antiquark confinement spectrum.
It is, however, well-known that higher-order contributions
to the meson propagator, in particular those from meson loops,
cannot be neglected.
Virtual meson loops give a correction to the meson mass, whereas decay
channels also contribute to the strong width of the meson.
One obtains for the propagator of a meson the form

\begin{equation}
\Pi(s)\;= \;\fnd{1}
{s\; -\;\left( M_\xrm{confinement}\; +\;\sum\;\Delta M_\xrm{meson loops}
\right)^{2}}
\;\;\; ,
\label{propag}
\end{equation}

\noindent
where $\Delta M$ develops complex values when open decay channels are involved.

For the full mass of a meson, all possible meson-meson loops have to be
considered.
A model for meson-meson scattering must therefore include all
possible inelastic channels as well.
Although in principle this could be done, in practice it is not feasible,
unless a scheme exists dealing with all vertices and their relative
intensities.
In Ref.~\cite{ZPC21p291} relative couplings have been determined in the
harmonic-oscillator approximation, assuming $^{3}P_{0}$ quark exchange.
However, further kinematical factors must be worked out and included.

In its present form, the model is far from perfect.
However, it allows for many predictions and for a good classification
of the mesonic resonances.
It moreover serves well as an interface between QCD quenched-lattice
calculations and experiment.
Quenched calculations describe confinement.
As we have shown, the confinement spectrum and wave functions are very
different from the hadronic reality for mesons.
Applying the model, we may indicate what differences must be anticipated.
For example, on the lattice one finds the lowest scalar-meson states at
1.3--1.5 GeV, exactly as we find for the pure harmonic oscillator.
But the model yields a further nonet of scalar resonances well below 1.0
GeV.
Hence, it is neither a failure of lattice QCD, nor of experiment!
The light scalar resonances just do not form part of the confinement spectrum.
\vspace{0.2cm}

{\bf The spectrum.}
The complete model consists of an expression for the $K$ matrix,
similar to formula (\ref{cotgdS}),
but extended to many meson-meson scattering channels,
several constituent quark-antiquark channels, and more
complicated transition potentials \cite{PRD27p1527,ZPC30p615},
which at the same time
and with the same set of four parameters reproduces bound states,
partial-wave scattering quantities, and the electromagnetic
transitions of $c\bar{c}$ and $b\bar{b}$ systems \cite{PRD44p2803}.

The $K$ matrix can be analytically continued below the various
thresholds, even the lowest one,
with no need of redefining any of the functions involved,
in order to study the singularities of the corresponding scattering matrix.
Below the lowest threshold, these poles show up on the real $\sqrt{s}$
axis, and can be interpreted as the bound states of the coupled system, to be
identified with the stable mesons.
For the light flavours one finds this way a nonet of light
pseudoscalars, i.e., the pion, Kaon, eta, and eta$'$.
For the heavy flavours, the lowest-lying model poles
can be identified with the $D(1870)$, $D_{s}(1970)$, $\eta_{c}(1S)$,
$J/\psi (1S)$, $\psi (3686)$, $B(5280)$, $B_{s}(5380)$,
$\Upsilon (1S)$, $\Upsilon (2S)$, and $\Upsilon (3S)$.

Above the lowest threshold, the model's partial-wave cross sections
and phase shifts for all included meson-meson channels
can be calculated and compared to experiment, as well as the
inelastic transitions.
Out of the many singularities of the scattering matrix in a rather complex
set of Riemann sheets, some come out with negative imaginary
part in the $\sqrt{s}$ plane, and moreover close enough to the physical real
axis so as to be noticed in the partial-wave phase shifts and cross sections.
These can be identified with the known resonances,
like the $\rho$ pole in $\pi\pi$ scattering,
or the $K^{\ast}$ pole in $K\pi$ scattering.
However, there may always be a pole in a nearby Riemann sheet
just around the corner of one of the thresholds, which can be
noticed in the partial-wave cross section.
The study of poles is an interesting subject by itself
\cite{NPB587p331,PRD59p074001}.
\vspace{0.2cm}

{\bf Scattering-matrix poles.}
In the hypothetical case of very small couplings for the three-meson
vertices, we obtain poles in the scattering matrix that are close
to the eigenvalues of the confinement spectrum.
Let us denote by $M_{1}$ and $M_{2}$ the meson masses,
and by $\Delta E$ the difference between the complex-energy pole
of the scattering matrix and the energy eigenvalue, $E_{N}$, of
the nearby state of the confinement spectrum.
Using formula (\ref{ERGR}), we obtain

\begin{equation}
\Delta E\;\approx\;
\left\{ R(s)\; -\; iI(s)\right\}\;\abs{{\cal F}_{N}}^{2}
\;\;\; .
\label{DeltaE}
\end{equation}

\noindent
We may distinguish two different cases:
\vspace{0.1cm}

(1) $E_{N}\;>\; M_{1}+M_{2}$ (above threshold),

(2) $E_{N}\;<\; M_{1}+M_{2}$ (below threshold).
\vspace{0.2cm}

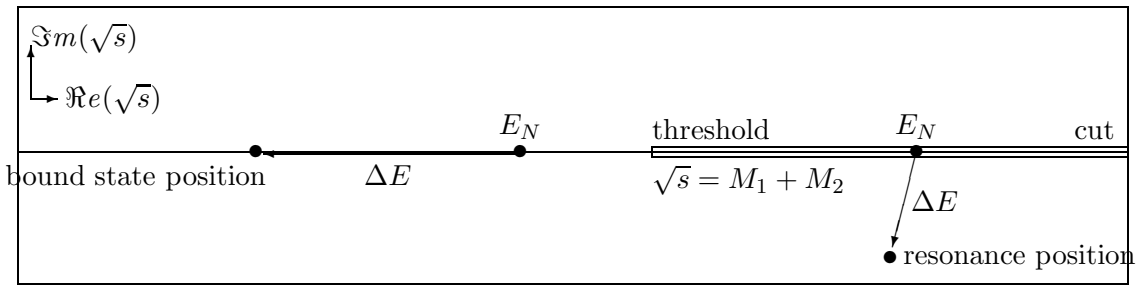
\begin{figure}[htbp]
\begin{center}
\begin{picture}(420,105)(0,-50)
\put(  0,-50){\line(1,0){420}}
\put(  0, 55){\line(1,0){420}}
\put(  0,-50){\line(0,1){105}}
\put(420,-50){\line(0,1){105}}
\put(0,0){\line(1,0){420}}
\put(240,-2){\line(0,1){4}}
\put(240,-2){\line(1,0){180}}
\put(240, 2){\line(1,0){180}}
\put(415,5){\makebox(0,0)[br]{\small cut}}
\put(240,5){\makebox(0,0)[bl]{\small threshold}}
\put(240,-5){\makebox(0,0)[tl]{\small $\sqrt{s}=M_{1}+M_{2}$}}
\put(340,0){\makebox(0,0){$\bullet$}}
\put(340,5){\makebox(0,0)[bc]{\small $E_{N}$}}
\put(340,0){\vector(-1,-4){9}}
\put(338,-15){\makebox(0,0)[lt]{\small $\Delta E$}}
\put(330,-40){\makebox(0,0){$\bullet$}}
\put(335,-40){\makebox(0,0)[lc]{\small resonance position}}
\put(190,0){\makebox(0,0){$\bullet$}}
\put(190,5){\makebox(0,0)[bc]{\small $E_{N}$}}
\put(190,-1){\vector(-1,0){97}}
\put(140,-5){\makebox(0,0)[tc]{\small $\Delta E$}}
\put(90,0){\makebox(0,0){$\bullet$}}
\put(94,-5){\makebox(0,0)[rt]{\small bound state position}}
\put(5,20){\vector(0,1){20}}
\put(5,43){\makebox(0,0)[cl]{\small $\Imag{\sqrt{s}}$}}
\put(5,20){\vector(1,0){10}}
\put(18,20){\makebox(0,0)[cl]{\small $\Real{\sqrt{s}}$}}
\end{picture}
\end{center}

\normalsize
\caption[]{When the confinement state on the real $\sqrt{s}$ axis is
below the lowest scattering threshold, then the bound-state singularity
comes out on the real $\sqrt{s}$ axis.
On the other hand, when the confinement state on the real $\sqrt{s}$ axis is in
the scattering continuum, then for small coupling (perturbative regime)
the resonance pole moves into the lower half of the complex $\sqrt{s}$ plane.}
\label{Ablow}
\end{figure}

When the nearby state of the confinement spectrum is in the
scattering continuum, then $\Delta E$ has a {\bf negative}
imaginary part and a real part, since both $R(s)$ and $I(s)$
of formula (\ref{DeltaE}) are real, and $I(s)$ is moreover positive.
The resonance singularity of the scattering matrix
is in the lower half of the complex-energy plane (second Riemann sheet),
as is depicted on the right-hand side of threshold in Fig.~(\ref{Ablow}).

When the nearby state of the confinement spectrum is below
the scattering threshold, then $\Delta E$ has only a real part,
since $I(s)$ turns purely imaginary below threshold, whereas
$R(s)$ remains real.
The bound-state singularity of the scattering matrix corresponding to
this situation remains on the real axis of the complex-energy plane,
as is depicted on the left-hand side of threshold in Fig.~(\ref{Ablow}).
\vspace{0.2cm}

{\bf Threshold behaviour.}
Near the lowest threshold, as a function of the overall coupling
constant, $S$-wave poles behave very differently
from $P$- and higher-wave poles.
This can easily be understood from the effective-range expansion
\cite{PotentialScattering} at the pole position.
There, the cotangent of the phase shift equals $i$.
Hence, for $S$ waves the next-to-lowest-order term in the expansion
equals $ik$ ($k$ represents the linear momentum related to $s$
and the lowest threshold).
For higher waves, on the other hand, the next-to-lowest-order term in the
effective-range expansion is proportional to $k^{2}$.

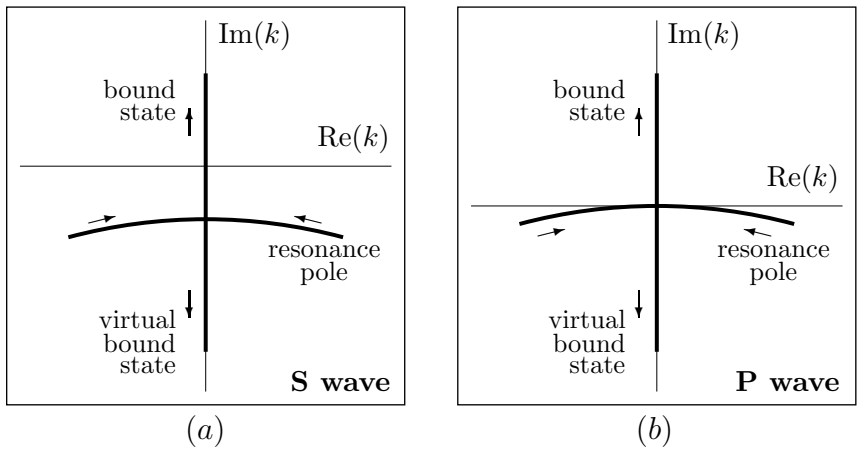
\begin{figure}[htbp]
\begin{center}
\begin{picture}(320,160)(0,-10)
\put(145,95){\makebox(0,0)[rb]{\small Re($k$)}}
\put(80,145){\makebox(0,0)[lt]{\small Im($k$)}}
\put(31,69){\vector(4,1){10}}
\put(119,69){\vector(-4,1){10}}
\put(69,102){\vector(0,1){10}}
\put(69,43){\vector(0,-1){10}}
\put(64,36){\makebox(0,0)[rt]{\footnotesize virtual}}
\put(64,27){\makebox(0,0)[rt]{\footnotesize bound}}
\put(64,18){\makebox(0,0)[rt]{\footnotesize state}}
\put(64,123){\makebox(0,0)[rt]{\footnotesize bound}}
\put(64,114){\makebox(0,0)[rt]{\footnotesize state}}
\put(120,60){\makebox(0,0)[ct]{\footnotesize resonance}}
\put(120,54){\makebox(0,0)[ct]{\footnotesize pole}}
\put(75,-4){\makebox(0,0)[ct]{$(a)$}}
\put(145,5){\makebox(0,0)[rb]{\small\bf S wave}}
\put(315,80){\makebox(0,0)[rb]{\small Re($k$)}}
\put(250,145){\makebox(0,0)[lt]{\small Im($k$)}}
\put(201,64){\vector(4,1){10}}
\put(289,64){\vector(-4,1){10}}
\put(239,102){\vector(0,1){10}}
\put(239,43){\vector(0,-1){10}}
\put(234,36){\makebox(0,0)[rt]{\footnotesize virtual}}
\put(234,27){\makebox(0,0)[rt]{\footnotesize bound}}
\put(234,18){\makebox(0,0)[rt]{\footnotesize state}}
\put(234,123){\makebox(0,0)[rt]{\footnotesize bound}}
\put(234,114){\makebox(0,0)[rt]{\footnotesize state}}
\put(290,60){\makebox(0,0)[ct]{\footnotesize resonance}}
\put(290,54){\makebox(0,0)[ct]{\footnotesize pole}}
\put(245,-4){\makebox(0,0)[ct]{$(b)$}}
\put(315,5){\makebox(0,0)[rb]{\small\bf P wave}}
\end{picture}
\end{center}

\normalsize
\caption[]{Variation of the positions of scattering-matrix poles
as a function of hypothetical variations in the three-meson-vertex coupling,
for $S$ waves ($a$), and for $P$ and higher waves ($b$).
The arrows indicate increasing coupling constant.}
\label{SPDpoles}
\end{figure}

Poles for $P$ and higher waves behave in the complex $k$ plane
as indicated in Fig.~(\ref{SPDpoles}$b$).
The two $k$-plane poles meet at threshold ($k=0$).
When the coupling constant of the model is increased, the poles
move along the imaginary $k$ axis.
One pole moves towards negative imaginary $k$, corresponding to
a virtual bound state below threshold on the real $\sqrt{s}$ axis,
but in the wrong Riemann sheet.
The other pole moves towards positive imaginary $k$,
corresponding to a real bound state.

For $S$-wave poles, the behaviour is shown in Fig.~(\ref{SPDpoles}$a$).
The two $k$-plane poles meet on the negative imaginary $k$ axis.
When the coupling constant of the model is slightly increased,
both poles continue on the negative imaginary $k$ axis,
corresponding to two virtual bound states below threshold on the real
$\sqrt{s}$ axis.
Upon further increasing the coupling constant of the model,
one pole moves towards increasing negative imaginary $k$, thereby
remaining a virtual bound state for all values of the coupling constant.
The other pole moves towards positive imaginary $k$,
eventually crossing threshold ($k=0$), thereby turning into a real
bound state of the system of coupled meson-meson scattering channels.
Hence, for a small range of hypothetical values of the coupling constant,
there are two virtual bound states, one of which is very close to
threshold.
Such a pole certainly has a noticeable influence on the scattering cross
section.
\clearpage

{\bf The low-lying nonet of S-wave poles.}
The nonet of low-lying $S$-wave poles behave as described above,
with respect to variations of the model's overall coupling constant.
However, they do not stem from the confinement spectrum,
but rather from the cavity.
For small values of the coupling, such poles disappear into the
continuum, i.e., they move towards negative imaginary infinity
\cite{HEPEX0106077},
and not towards an eigenstate of the confinement spectrum as in
Fig.~(\ref{Ablow}).

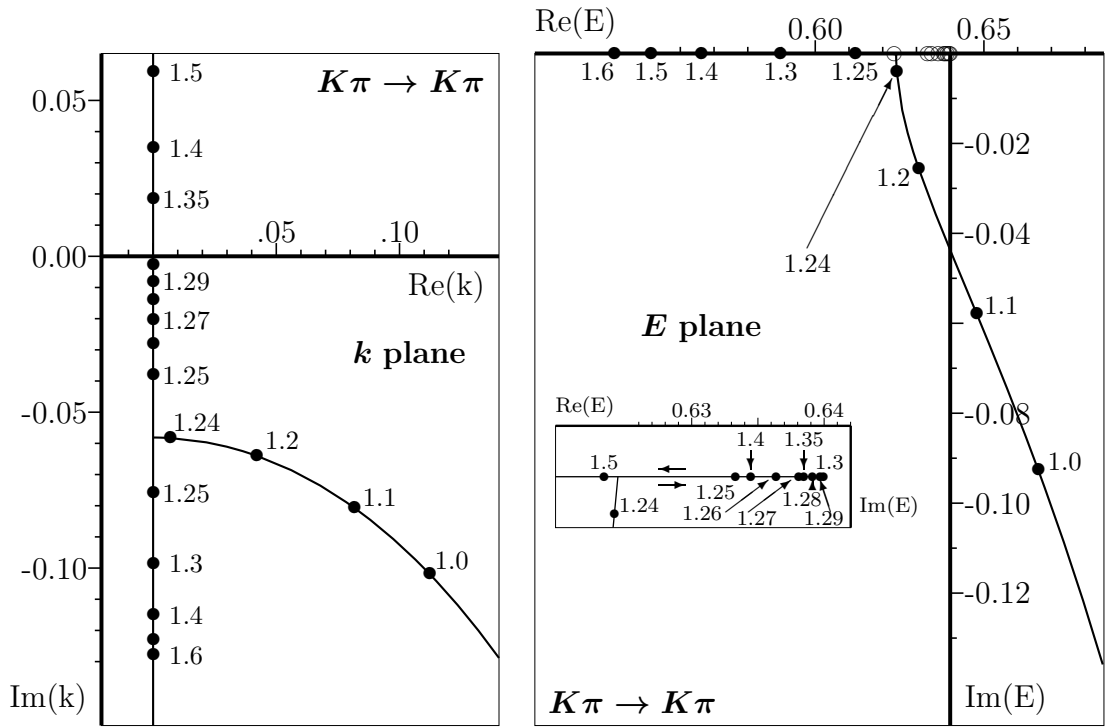
\begin{figure}[htbp]
\begin{center}
\begin{picture}(190,280)(-30,0)
\put(66.11,183.15){\makebox(0,0)[bc]{.05}}
\put(112.71,183.15){\makebox(0,0)[bc]{.10}}
\put(-5.52,59.61){\makebox(0,0)[rc]{-0.10}}
\put(-5.52,118.62){\makebox(0,0)[rc]{-0.05}}
\put(-5.52,177.63){\makebox(0,0)[rc]{0.00}}
\put(-5.52,236.64){\makebox(0,0)[rc]{0.05}}
\put(145,172){\makebox(0,0)[tr]{Re(k)}}
\put(-5.52,4.01){\makebox(0,0)[br]{Im(k)}}
\put(124.07,57.41){\makebox(0,0){$\bullet$}}
\put(95.66,82.38){\makebox(0,0){$\bullet$}}
\put(58.69,102.13){\makebox(0,0){$\bullet$}}
\put(26.13,108.83){\makebox(0,0){$\bullet$}}
\put(19.53,88.07){\makebox(0,0){$\bullet$}}
\put(19.53,61.34){\makebox(0,0){$\bullet$}}
\put(19.53,41.77){\makebox(0,0){$\bullet$}}
\put(19.53,32.35){\makebox(0,0){$\bullet$}}
\put(19.53,26.76){\makebox(0,0){$\bullet$}}
\put(19.53,247.47){\makebox(0,0){$\bullet$}}
\put(19.53,218.55){\makebox(0,0){$\bullet$}}
\put(19.53,199.28){\makebox(0,0){$\bullet$}}
\put(19.53,174.22){\makebox(0,0){$\bullet$}}
\put(19.53,168.04){\makebox(0,0){$\bullet$}}
\put(19.53,161.23){\makebox(0,0){$\bullet$}}
\put(19.53,153.53){\makebox(0,0){$\bullet$}}
\put(19.53,144.48){\makebox(0,0){$\bullet$}}
\put(19.53,132.76){\makebox(0,0){$\bullet$}}
\footnotesize
\put(132.40,62.41){\makebox(0,0){1.0}}
\put(105.23,86.38){\makebox(0,0){1.1}}
\put(68.26,108.13){\makebox(0,0){1.2}}
\put(35.99,114.83){\makebox(0,0){1.24}}
\put(32,88.07){\makebox(0,0){1.25}}
\put(32,61.34){\makebox(0,0){1.3}}
\put(32,41.77){\makebox(0,0){1.4}}
\put(32,26.76){\makebox(0,0){1.6}}
\put(32,247.47){\makebox(0,0){1.5}}
\put(32,218.55){\makebox(0,0){1.4}}
\put(32,199.28){\makebox(0,0){1.35}}
\put(32,168.04){\makebox(0,0){1.29}}
\put(32,153.53){\makebox(0,0){1.27}}
\put(32,132.76){\makebox(0,0){1.25}}
\normalsize
\put(145,248){\makebox(0,0)[tr]{\bm{K\pi\rightarrow K\pi}}}
\put(95,135){\makebox(0,0)[bl]{\bm{k} {\bf plane}}}
\end{picture}
\begin{picture}(215,280)(0,0)
\put(106.06,259.94){\makebox(0,0)[bc]{0.60}}
\put(169.87,259.94){\makebox(0,0)[bc]{0.65}}
\put(162.63,50.13){\makebox(0,0)[lc]{-0.12}}
\put(162.63,84.18){\makebox(0,0)[lc]{-0.10}}
\put(162.63,118.23){\makebox(0,0)[lc]{-0.08}}
\put(162.63,186.32){\makebox(0,0)[lc]{-0.04}}
\put(162.63,220.37){\makebox(0,0)[lc]{-0.02}}
\put(0.00,259.94){\makebox(0,0)[bl]{Re(E)}}
\put(162.63,4.01){\makebox(0,0)[bl]{Im(E)}}
\put(190.44,96.66){\makebox(0,0){$\bullet$}}
\put(167.12,155.88){\makebox(0,0){$\bullet$}}
\put(145.21,210.79){\makebox(0,0){$\bullet$}}
\put(136.90,247.46){\makebox(0,0){$\bullet$}}
\put(121.19,254.42){\makebox(0,0){$\bullet$}}
\put(92.86,254.42){\makebox(0,0){$\bullet$}}
\put(62.99,254.42){\makebox(0,0){$\bullet$}}
\put(43.91,254.42){\makebox(0,0){$\bullet$}}
\put(30.10,254.42){\makebox(0,0){$\bullet$}}
\scriptsize
\put(135.88,254.42){\makebox(0,0){$\odot$}}
\put(150.01,254.42){\makebox(0,0){$\odot$}}
\put(155.14,254.42){\makebox(0,0){$\odot$}}
\put(157.06,254.42){\makebox(0,0){$\odot$}}
\put(156.72,254.42){\makebox(0,0){$\odot$}}
\put(155.98,254.42){\makebox(0,0){$\odot$}}
\put(154.66,254.42){\makebox(0,0){$\odot$}}
\put(152.46,254.42){\makebox(0,0){$\odot$}}
\put(148.54,254.42){\makebox(0,0){$\odot$}}
\footnotesize
\put(194,96.66){\makebox(0,0)[bl]{1.0}}
\put(170,155.88){\makebox(0,0)[bl]{1.1}}
\put(142,210.79){\makebox(0,0)[tr]{1.2}}
\put(103,181){\vector(1,2){32}}
\put(103,178){\makebox(0,0)[tc]{1.24}}
\put(121,251){\makebox(0,0)[tc]{1.25}}
\put(93,251){\makebox(0,0)[tc]{1.3}}
\put(63,251){\makebox(0,0)[tc]{1.4}}
\put(44,251){\makebox(0,0)[tc]{1.5}}
\put(30,251){\makebox(0,0)[tr]{1.6}}
\scriptsize
\put(5,75){\makebox(0,0)[bl]{
\begin{picture}(141.73,66.69)(0.00,0.00)
\put(51.51,41.90){\makebox(0,0)[bc]{0.63}}
\put(101.59,41.90){\makebox(0,0)[bc]{0.64}}
\put(0.00,41.90){\makebox(0,0)[bl]{Re(E)}}
\put(115.02,4.01){\makebox(0,0)[bl]{Im(E)}}
\put(39,16.2){\vector(1,0){10}}
\put(49,22.2){\vector(-1,0){10}}
\put(22.16,5){\makebox(0,0){$\bullet$}}
\put(18.29,19.27){\makebox(0,0){$\bullet$}}
\put(73.72,19.27){\makebox(0,0){$\bullet$}}
\put(93.85,19.27){\makebox(0,0){$\bullet$}}
\put(101.39,19.27){\makebox(0,0){$\bullet$}}
\put(100.07,19.27){\makebox(0,0){$\bullet$}}
\put(97.15,19.27){\makebox(0,0){$\bullet$}}
\put(92.00,19.27){\makebox(0,0){$\bullet$}}
\put(83.37,19.27){\makebox(0,0){$\bullet$}}
\put(67.98,19.27){\makebox(0,0){$\bullet$}}
\put(24,6){\makebox(0,0)[bl]{1.24}}
\put(18.3,22){\makebox(0,0)[bc]{1.5}}
\put(73.7,29){\vector(0,-1){7}}
\put(73.7,31){\makebox(0,0)[bc]{1.4}}
\put(93.9,29){\vector(0,-1){7}}
\put(93.9,31){\makebox(0,0)[bc]{1.35}}
\put(109,22){\makebox(0,0)[br]{1.3}}
\put(104,6){\vector(-1,3){4}}
\put(109,6){\makebox(0,0)[tr]{1.29}}
\put(97.2,15){\vector(0,1){3}}
\put(93.2,13.5){\makebox(0,0)[tc]{1.28}}
\put(73,6){\vector(4,3){16}}
\put(76,6){\makebox(0,0)[tc]{1.27}}
\put(64.4,6){\vector(4,3){16}}
\put(63,8){\makebox(0,0)[tr]{1.26}}
\put(68,16){\makebox(0,0)[tr]{1.25}}
\end{picture}
}}
\normalsize
\put(5,5){\makebox(0,0)[bl]{\bm{K\pi\rightarrow K\pi}}}
\put(40,145){\makebox(0,0)[bl]{\bm{E} {\bf plane}}}
\end{picture}
\end{center}

\normalsize
\caption[]{Hypothetical movement of the $K_{0}^{\ast}(730)$ pole in
$K\pi$ $S$-wave scattering as a function of the coupling constant.
The two branches on the imaginary $k$ axis are discussed in the text.
In the $E=\sqrt{s}$ plane these two branches come out on the real axis
below threshold.
The poles of the upper branch are shown as open circles in the main
figure, and as closed circles in the inset.
Units are in GeVs.}
\label{kappapole}
\end{figure}

In Fig.~(\ref{kappapole}) we study the hypothetical pole
positions of the $K_{0}^{\ast}(730)$ pole in $K\pi$
$S$-wave scattering.
The physical value of the coupling constant equals 0.75,
which is not shown in Fig.~(\ref{kappapole}).
A figure for smaller values of the coupling constants can be
found in Ref.~\cite{HEPEX0106077}.
The physical pole in $K\pi$ isodoublet $S$-wave scattering,
related to experiment \cite{HEPEX0204018,HEPEX0110052},
comes out at $727-i263$ MeV in Ref.~\cite{ZPC30p615}.
Here we concentrate on the threshold behaviour of the
hypothetical pole movements in the complex $k$ and $\sqrt{s}$
planes.
Until they meet on the axis, which is for a value of the
coupling constant slightly larger than 1.24, we have only
depicted the right-hand branch.

In the left-hand picture of Fig.~(\ref{kappapole}) we observe how the poles
arrive on the imaginary $k$ axis, and then continue to move along that axis.
One of the poles moves upwards, initially describing a virtual
bound state, and crossing the real $k$ axis for a value of the
coupling constant slightly larger than 1.30.
The other pole moves downwards, remaining a virtual bound state
for further increasing values of the coupling constant.

In the right-hand picture of Fig.~(\ref{kappapole}) the same pole has been
depicted in the complex $\sqrt{s}$ plane.
Here, the situation is more confusing, since the pole positions
are in the same interval of energies.
The pole corresponding to the one moving downwards
along the imaginary $k$ axis moves to the left on the real
$\sqrt{s}$ axis.
Its positions as a function of the coupling constant are
indicated by solid circles.
The pole moving upwards along the imaginary $k$ axis
initially moves towards threshold and then turns back,
following the former pole, but in a different Riemann sheet.
The positions of the latter pole are indicated by open circles.
In the inset we try to better clarify its motion.
Notice that, since we took 0.14 GeV and 0.50 GeV for
the pion and the Kaon mass, respectively, we end up
with a threshold at 0.64 GeV.

It is interesting to notice that in a recent work of Boglione and
Pennington \cite{HEPPH0203149} also a zero-width state is found
below the $K\pi$ threshold in $S$-wave scattering.
Here, we obtain such a state for \em unphysical \em \/values of the coupling.

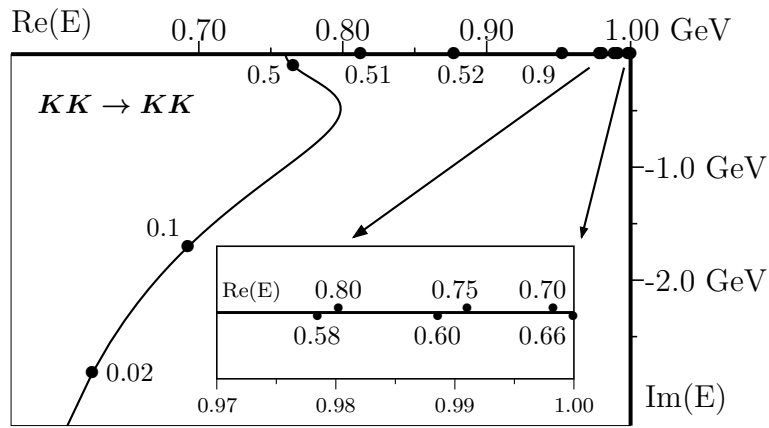
\begin{figure}[htbp]
\begin{center}
\begin{picture}(283.46,180.08)(0.00,0.00)
\put(71.03,146.12){\makebox(0,0)[bc]{0.70}}
\put(125.46,146.12){\makebox(0,0)[bc]{0.80}}
\put(179.90,146.12){\makebox(0,0)[bc]{0.90}}
\put(224.34,146.12){\makebox(0,0)[bl]{1.00 GeV}}
\put(239.86,55.06){\makebox(0,0)[lc]{-2.0 GeV}}
\put(239.86,97.83){\makebox(0,0)[lc]{-1.0 GeV}}
\put(0.00,146.12){\makebox(0,0)[bl]{Re(E)}}
\put(239.86,4.01){\makebox(0,0)[bl]{Im(E)}}
\put(132.10,140.60){\makebox(0,0){$\bullet$}}
\put(167.43,140.60){\makebox(0,0){$\bullet$}}
\put(222.60,140.60){\makebox(0,0){$\bullet$}}
\put(228.10,140.60){\makebox(0,0){$\bullet$}}
\put(234.30,140.60){\makebox(0,0){$\bullet$}}
\put(233.39,140.60){\makebox(0,0){$\bullet$}}
\put(229.44,140.60){\makebox(0,0){$\bullet$}}
\put(223.56,140.60){\makebox(0,0){$\bullet$}}
\put(208.43,140.60){\makebox(0,0){$\bullet$}}
\put(30.65,19.70){\makebox(0,0){$\bullet$}}
\put(66.78,67.54){\makebox(0,0){$\bullet$}}
\put(106.72,136.06){\makebox(0,0){$\bullet$}}
\footnotesize
\put(144,136){\makebox(0,0)[rt]{0.51}}
\put(179,136){\makebox(0,0)[rt]{0.52}}
\put(206,136){\makebox(0,0)[rt]{0.9}}
\put(36,20){\makebox(0,0)[lc]{0.02}}
\put(64,75){\makebox(0,0)[rc]{0.1}}
\put(102,136){\makebox(0,0)[rt]{0.5}}
\put(10,125){\makebox(0,0)[lt]{\bm{KK\rightarrow KK}}}
\scriptsize
\put(75,33){\makebox(0,0)[bl]{
\begin{picture}(170,60)(0,138.78)
\put(0.17,116){\makebox(0,0)[tc]{0.97}}
\put(45.03,116){\makebox(0,0)[tc]{0.98}}
\put(90.06,116){\makebox(0,0)[tc]{0.99}}
\put(135.03,116){\makebox(0,0)[tc]{1.00}}
\put(2.00,152.26){\makebox(0,0)[bl]{Re(E)}}
\put(37.94,147.34){\makebox(0,0){$\bullet$}}
\put(83.46,147.34){\makebox(0,0){$\bullet$}}
\put(134.76,147.34){\makebox(0,0){$\bullet$}}
\put(127.26,150.34){\makebox(0,0){$\bullet$}}
\put(94.58,150.34){\makebox(0,0){$\bullet$}}
\put(45.91,150.34){\makebox(0,0){$\bullet$}}
\footnotesize
\put(37.94,143){\makebox(0,0)[tc]{0.58}}
\put(83.46,143){\makebox(0,0)[tc]{0.60}}
\put(132,143){\makebox(0,0)[tr]{0.66}}
\put(132,154){\makebox(0,0)[br]{0.70}}
\put(90,154){\makebox(0,0)[bc]{0.75}}
\put(45.91,154){\makebox(0,0)[bc]{0.80}}
\end{picture}}}
\end{picture}
\end{center}
\normalsize
\caption[]{Pole movement as a function of the coupling constant for
$KK$ $I=1$ $S$-wave scattering.
Some values of the coupling constant are indicated in the figure.
The six filled circles at the right end of the real axis correspond, from
left to right, to the values 0.58, 0.80, 0.60, 0.75, 0.70, and 0.66
for the model's coupling constant. This situation is magnified in the inset.}
\label{a0pole}
\end{figure}

In Fig.~(\ref{a0pole}) we have depicted the movement of the
$a_ {0}(980)$ pole in $S$ wave $I=1$ $KK$ scattering
(threshold at 1.0 GeV) on the upwards-going branch.
One observes a very similar behaviour as in the case of $K\pi$
scattering, but with two important differences, to be described next.

The $K_{0}^{\ast}(730)$ poles meet on the real $\sqrt{s}$ axis
only 16 MeV below threshold (see Fig.~\ref{kappapole}),
and for a value of the coupling constant which is well above
the physical value of 0.75, whereas the $a_ {0}(980)$ poles meet
238 MeV below threshold, when the coupling constant
only equals 0.51.
At the physical value of the coupling constant, the $a_ {0}(980)$
pole is a real bound state some 9 MeV below threshold.

But there is yet another difference.
Whereas the $K\pi$ channel represents the lowest possible scattering
threshold for the $K_{0}^{\ast}(730)$ system,
$KK$ is not the lowest channel for the $a_ {0}(980)$.
In a more complete description, at least all pseudoscalar
meson-meson loops should be taken into account.
One of these is the $\eta\pi$ channel, which has a threshold
well below $KK$.
Consequently, upon including the $\eta\pi$ channel in the model, the pole
cannot remain on the real $\sqrt{s}$ axis, but has to acquire
an imaginary part, in a similar way as shown in
Fig.~(\ref{Ablow}).
In Ref.~\cite{ZPC30p615} we obtained a resonance-like structure
in the $\eta\pi$ cross section, representing the physical
$a_{0}(980)$. The corresponding pole came out at $962-i28$ MeV.

For the $f_{0}(980)$ system the situation is very similar
to that of the $a_ {0}(980)$.
Assuming a pure $s\bar{s}$ quark content \cite{PLB521p15}, we obtain for the
variation of the corresponding pole in $KK$ $I=0$ $S$-wave scattering
a picture almost equal to the one shown in Fig.~(\ref{a0pole}).
However, only in lowest order the $KK$ channel could be considered
the lowest threshold for the $f_{0}(980)$ system.
In reality, $s\bar{s}$ also couples to the non-strange quark-antiquark
isosinglet through $KK$, and hence to $\pi\pi$ \cite{NPB266p451}.
This coupling is nevertheless very weak, which implies that the
resulting pole does not move far away from the $KK$ bound state.
In Ref.~\cite{ZPC30p615} we obtained a resonance-like structure
in the $\pi\pi$ cross section representing the physical
$f_ {0}(980)$.
The corresponding pole came out at $994-i20$ MeV.

At lower energies, we found for the same cross section a pole
which is the equivalent of the $K_{0}^{\ast}(730)$ system, but now
in $\pi\pi$ isoscalar $S$-wave scattering.
This pole at $470-i208$ MeV may be associated with the
$\sigma$ meson, since it has the same quantum numbers,
and lies in the ballpark of predicted pole positions
in models of the $\sigma$ (for a complete overview of $\sigma$ poles,
see Ref.~\cite{HEPPH0201006}).

We do not find any other relevant poles in the energy region
up to 1.0 GeV.
\vspace{0.2cm}

{\bf Conclusions.}
We have shown that the poles of the $a_{0}(980)$ and $f_{0}(980)$
belong to a nonet of scattering-matrix poles.
The lower-lying isoscalar pole and the isodoublet poles in the complex-energy
plane have real parts of 0.47 GeV and 0.73 GeV, respectively, and imaginary
parts of 0.21 GeV resp.\ 0.26 GeV.
Whether these poles represent real physical resonances
\cite{NPA688p823} is not so relevant here.
Important is that the $a_{0}(980)$ and $f_{0}(980)$ are well
classified within a nonet of scattering-matrix poles with very
specific characteristics, different from those of the poles stemming
from confinement, like the confinement-ground-state nonet of scalar mesons
$f_{0}(1370)$, $a_{0}(1450)$, $K_{0}^{\ast}(1430)$, and $f_{0}(1500)$.
The latter poles vary as a function of the coupling constant
exactly the way indicated in figure (\ref{Ablow}).
For vanishing coupling constant they end up on the real $\sqrt{s}$ axis
at the positions of the various ground-state eigenvalues
of the confinement spectrum, which are the light-flavour
$^{3}P_{0}$ states at 1.3 to 1.5 GeV
\cite{PRD61p014015,NPPS53p236}.

The low-lying $S$-wave poles related to the cross sections
in the $f_{0}(470)$, $K_{0}^{\ast}(730)$, $f_{0}(980)$, and $a_{0}(980)$
regions move to negative imaginary infinity in the $\sqrt{s}$ plane
for decreasing values of the coupling.
The pole positions for the physical value of the coupling are well
explained by their threshold behaviour.
Whether or not these poles have large imaginary parts, leading to large
widths and strong resonance distortion, depends in a subtle way on the
thresholds and couplings of the various relevant scattering channels
\cite{PLB462p14}.

As to the nature of the light scalar mesons, which has recently been discussed
in Refs.~\cite{HEPPH0204205,HEPPH0201171,NPA675p209c,PRD65p114011},
we can only remark that in a many-coupled-channel model each of the
channels contributes to the states under the resonance, not just one
specific channel.

We have moreover shown that the
{\bf principle of flavour independence for the strong interactions}
has far-reaching consequences for the construction of hadron models.
Applied as a guiding principle for building a largely non-relativistic
many-coupled-channel Schr\"{o}dinger equation for meson-meson scattering,
it results in a surprisingly complete model for mesons and mesonic
resonances.
Its description of the light scalar resonances leaves no doubt on where
these mesons are, nor on how to classify them.
\vspace{0.3cm}

{\bf Acknowledgement}:
This work has been partly supported by the
{\it Funda\c{c}\~{a}o para a Ci\^{e}ncia e a Tecnologia}
of the {\it Minist\'{e}rio da
Ci\^{e}ncia e da Tecnologia} \/of Portugal,
under contract numbers
POCTI/\-35304/\-FIS/\-2000,
CERN/\-FIS/\-43697/\-2001,
PRAXIS XXI/\-BPD/\-20186/\-99,
and
SFRH/\-BPD/\-9480/\-2002.

\end{document}